\documentclass[aps,prd,preprint,groupedaddress]{revtex4-1}
\usepackage[utf8]{inputenc}
\usepackage[english]{babel}
\usepackage[letterpaper,margin=1in]{geometry}

\usepackage[dvipsnames]{xcolor}
\usepackage{amsmath,amssymb,amsfonts,enumerate,float,graphicx,hyperref,mathrsfs,mathtools,fancyvrb,slashed,pdfpages,empheq,tabularx,caption,multirow,subcaption,graphics}

\usepackage[inline,shortlabels]{enumitem}

\allowdisplaybreaks

\usepackage{colortbl}
\usepackage{tikz,pgfplots}

\usetikzlibrary{calc,patterns,decorations.pathmorphing,decorations.markings,arrows}

\graphicspath{{./figures/}}

\usepackage[most]{tcolorbox}

\tcbset{
	frame code={}
	center title,
	left=0pt,
	right=0pt,
	top=0pt,
	bottom=0pt,
	colback=yellow!25,
	colframe=white,
	width=\dimexpr\textwidth\relax,
	enlarge left by=0mm,
	boxsep=5pt,
	arc=0pt,outer arc=0pt,
}

\newcommand\myshade{85}
\hypersetup{
	linkcolor  = violet!\myshade!black,
	citecolor  = YellowOrange!\myshade!black,
	urlcolor   = Aquamarine!\myshade!black,
	colorlinks = true,
}

\newcommand{\pp}[1]{\left ( #1 \right )}

\newcommand{\vv}[1]{\left \vert #1 \right \vert}

\newcommand{\lag}{\mathscr L}

\newcommand{\nn}{\nonumber\\ &}

\newcommand{\ar}{\nonumber \\ & +}

\makeatletter
\AtBeginDocument{\let\LS@rot\@undefined}
\makeatother

\begin{document}
\title{Strong $ B _{QQ'}^* B_{QQ'} V $ vertices and the radiative decays of $ B _{QQ}^* \to  B_{QQ} \gamma $ in the light-cone sum rules}

\author{T. M. Aliev}
\email[]{taliev@metu.edu.tr}
\affiliation{Physics Department, Middle East Technical University, Ankara 06800, Turkey}

\author{T. Barakat}
\email[]{tbarakat@ksu.edu.sa}
\affiliation{Physics \& Astronomy Department, King Saud University, Riyadh 11451, Saudi Arabia}

\author{K. \c Sim\c sek}
\email[]{ksimsek@u.northwestern.edu}
\affiliation{Department of Physics \& Astronomy, Northwestern University, Evanston, Illinois 60208, USA}

\date{\today}

\begin{abstract}
	The strong coupling constants of spin-3/2 to spin-1/2 doubly heavy baryon transitions with light vector mesons are estimated within the light-cone QCD sum rules method. Moreover, using the vector-meson dominance ans\"atz, the widths of radiative decays $ B _{QQ}^* \to B _{QQ} \gamma $ are calculated. The results for the said decay widths are compared to the predictions of other approaches. 
\end{abstract}
\maketitle 
\section{Introduction}\label{sec:1}
The quark model is a vital tool for the classification of hadronic states. %Many states predicted by the quark model have already been observed in experiments. 
It predicts the existence of numerous doubly heavy baryons. Among various doubly heavy baryon states, only two, namely $ \Xi _{cc}^{++} $ and $ \Xi _{cc}^+ $, have been observed. The first observation of $ \Xi_{cc}^+ $ was announced by the SELEX Collaboration in the channels $ \Xi _{cc}^{+} \to \Lambda _c^+ K^- \pi ^+ $ and $ pD^+K^- $ with a mass $ 3518.7 \pm 1.7 {\rm\ MeV} $ \cite{Mattson2002}. In 2017, the LHCb Collaboration announced an observation of the doubly heavy baryon $ \Xi _{cc}^{++} $ in the mass spectrum $ \Lambda _c^+ K^- \pi^+ \pi^- $ \cite{Aaij2017} and confirmed also by measuring another decay channel, $ \Xi _{cc}^{++} \to \Xi _c^+ \pi^+ $ \cite{Aaij2018}, with an average mass obtained as $ 3621.24 \pm 0.65 \pm 0.31 {\rm\ MeV} $. The observation of doubly heavy baryon states stimulated new experimental studies in this direction \cite{Aaij2019,Aaij2020}.
\par 
Theoretical studies on this subject include the study of weak, electromagnetic, and strong decays of doubly heavy baryons. Their weak and strong decays have been comprehensively analyzed within the framework of the light-front QCD, QCD sum rules, and the light-cone sum rules (LCSR) method \cite{Wang2017, Xiao2017, Wang2017b, Cheng2018, Shi2018, Zhao2018, Shi2020, Shi2019, Hu2020}. Their electromagnetic properties and radiative decays have been discussed in \cite{Li2017, Meng2017, Li2017b, Bahtiyar2018}. The strong couplings of doubly heavy baryons with light mesons within the light-cone sum rules have been studied in \cite{Rostami2020, Aliev2020, Aliev2020b, Alrebdi2020, Azizi2020}. These coupling constants are the main parameters for understanding the dynamics of strong decays. The coupling constants of spin-3/2 to spin-1/2 doubly heavy baryons with $ \rho^+ $ and $ K^* $ have been studied in \cite{Aliev2020} within the framework of the LCSR method.
\par 
The aim of this work is two-fold. First, we extend our previous work \cite{Aliev2020} to study the vertices $ \Xi _{QQ'q}^*  \Xi _{QQ'q}\omega $ and $ \Omega _{QQ's}^*  \Omega _{QQ's} \phi $, where $ \Xi _{QQ'q}^* $ ($ \Omega_{QQ's}^* $) and $ \Xi _{QQ'q} $ ($ \Omega _{QQ's} $) denote the spin-3/2 and spin-1/2 doubly heavy baryons, respectively, within the LCSR method and second, using the results for these vertices and assuming the vector-meson dominance (VMD), we estimate the radiative decay widths of $ \Xi _{QQq}^* \to \Xi _{QQq} \gamma $ and $ \Omega _{QQs}^* \to \Omega _{QQs} \gamma $. In all the following discussion, we will denote the spin-3/2 (1/2) doubly heavy baryons by $ B _{QQ'}^* $ ($ B _{QQ'} $) customarily.
\par 
The paper is organized as follows. In Sec. \ref{sec:2}, first, we derive the LCSR for the coupling constants of the light vector mesons $ \omega $ and $ \phi $ for the $ \Xi _{QQ'q}^* \Xi _{QQ'q}\omega $ and $ \Omega _{QQ's}^* \Omega _{QQ's}\phi $ vertices; second, we present the results for the radiative decays $ \Xi _{QQq}^*\to \Xi _{QQq}\gamma $ and $ \Omega _{QQs}^* \to \Omega _{QQs}\gamma $ by assuming the VMD. Sec. \ref{sec:3} contains the numerical analysis of the obtained sum rules for the strong coupling constants and radiative decays. A summary and conclusion are presented in Sec. \ref{sec:4}.
\section{The $ B _{QQ'q}^* B _{QQ'q} V $ vertices in the light-cone sum rules}\label{sec:2}
By using the Lorentz invariance, the vertices $ B _{QQ'}^* B_{QQ'}V $, where $ V = \rho^0 $, $ \omega $, or $ \phi $ and $ B^{(*)} = \Xi^{(*)} $ or $ \Omega^{(*)} $, are parametrized in terms of three coupling constants, $ g_1 $, $ g_2 $, and $ g_3 $, as follows \cite{Jones1973}:
\begin{align}
	\langle V (q) B _{QQ'}^*(p_2) \vert B _{QQ'}(p_1) \rangle &= \bar u _\alpha (p_2) [
		g_1 (\varepsilon ^{*\alpha} \slashed q - q ^\alpha \slashed \varepsilon^*) \gamma _5
		+ g_2 (P\cdot q \varepsilon ^{*\alpha} - P \cdot \varepsilon^* q^\alpha)\gamma _5 
		\ar g_3 (q\cdot \varepsilon^* q^\alpha - q^2 \varepsilon ^{*\alpha}) \gamma _5		
	] u(p_1) \label{1}
\end{align}
where $ u_\alpha (p_2) $ is the Rarita-Schwinger spinor for a spin-3/2 baryon, $ \varepsilon _\alpha $ is the 4-polarization vector of the light vector meson $ V $, $ P = (p_1+p_2)/2 $, and $ q=p_1-p_2 $. In the rest of the text, we denote $ p_2=p $ and $ p_1=p+q $. 
\par 
For the determination of the said three coupling constants, $ g_1 $, $ g_2 $, and $ g_3 $, within the LCSR, we introduce the following correlation function:
\begin{align}
	\Pi _\mu (p,q) = i \int d^4x\ e^{ipx} \langle V(q) \vert \mathrm T\{\eta _\mu (x) \bar \eta (0)\} \vert 0 \rangle 
\end{align}
where $ V(q) $ is a light vector meson ($ \rho^0 $, $ \omega $, or $ \phi $) with 4-momentum $ q_\mu $, and $ \eta _\mu $ and $ \eta $ are the interpolating currents for the spin-3/2 and spin-1/2 baryons, respectively. The most general form of the interpolating currents of spin-3/2 and spin-1/2 baryons doubly heavy baryons are
\def\T{{\rm T}}
\begin{align}
	\eta _\mu &= N \epsilon^{abc} \{
		({q^a}^\T C \gamma _\mu Q^b) Q'^c 
		+ ({q^a}^\T C \gamma _\mu Q'^b) Q^c
		+ ({Q^a}^\T C \gamma _\mu Q'^b) q^c
	\}\\
	\eta ^{(S)} &= \frac 1{\sqrt 2}\epsilon^{abc} \sum _{i=1}^2 [
		({Q^a}^\T A_1^i q^b) A_2^i Q'^c
		+ (Q \leftrightarrow Q')
	]\\
	\eta ^{(A)} &= \frac 1{\sqrt 6} \epsilon^{abc} \sum _{i=1}^2 [
		2(Q^a A_1^i Q'^b)A_2^i q^c
		+({Q^a}^\T A_1^i Q'^c) 
		- {Q'^a}^\T A_1^i q^b
	] A_2^i Q^c
\end{align}
where $ \mathrm T $ is the transpose, $ N = \sqrt{1/3} $ ($ \sqrt{2/3} $) for identical (distinct) heavy quarks, $ A_1^1 = C $, $ A_2^1 = \gamma _5 $, $ A_1^2 = C \gamma _5 $, and $ A_2^2 = \beta I $, the superscripts $ S $ and $ A $ denote symmetric and antisymmetric interpolating currents with respect to the interchange of heavy quarks, and $ \beta $ is the arbitrary parameter, for which $ \beta = -1 $ corresponds to the case of the Ioffe current. 
\par 
The LCSR for the coupling constants, $ g_1 $, $ g_2 $, and $ g_3 $, is obtained by calculating the correlation function in two different regions: First, in terms of hadrons, and second, in the deep Euclidean domain by using operator product expansion (OPE). In terms of hadrons, the correlation function is obtained by inserting a complete set of intermediate hadronic states carrying the same quantum numbers as the interpolating currents $ \eta _\mu $ and $ \eta $ and using the quark-hadron duality. After isolating the ground state contribution, we get
\begin{align}
	\Pi _\mu (p,q) &= \frac{\lambda _1 \lambda _2} {(m_1^2 - p^2)[m_2^2 - (p+q)^2]} [
		-g_1 (m_1+m_2) \slashed \varepsilon^* \slashed p \gamma _5 q _\mu
		+g_2 \slashed q \slashed p \gamma _5 p \cdot \varepsilon^* q_\mu 
		+g_3 q^2 \slashed q \slashed p\gamma _5 \varepsilon _\mu^*
		\ar {\rm other\ structures}
	] \label{4}
\end{align}
Here, $\varepsilon^\mu$ is the 4-polarization vector of the light vector meson. In the derivation of Eq. \eqref{4}, the following definitions have been used:
\begin{gather}
	\langle 0 \vert \eta \vert B _{QQ'} (p) \rangle = \lambda _1 u (p,s)\\
	\langle 0 \vert \eta _\mu \vert B _{QQ'}^* (p) \rangle = \lambda _2 u_\mu (p,s)
\end{gather}
where $\lambda_1$ ($m_1$) and $\lambda_2$ ($m_2$) are the residues (masses) of the spin-3/2 and spin-1/2 states, respectively. The summation over spin-1/2 and spin-3/2 baryons is performed by using the corresponding completeness relations:
\begin{gather}
	\sum _s u(p,s) \bar u(p,s) = \slashed p + m\\
	\sum _s u _\mu (p,s) \bar u _\nu (p,s) = -(\slashed p + m) \Big[
		g _{\mu\nu} - \frac 13 \gamma _\mu \gamma _\nu - \frac 23 \frac{p_\mu p_\nu}{m^2} +\frac 13 \frac{p_\mu \gamma_\nu - p_\nu \gamma _\mu} m
	\Big] \label{6.2}
\end{gather}
At this point, we would like to make the following remarks:
\begin{enumerate}
	[(a)]
	\item The current $ \eta _\mu $ couples also to spin-1/2 baryons, $ B(p) $, with the corresponding matrix element
	\begin{align}
		\langle 0 \vert \eta _\mu \vert B^- (p) \rangle = A\pp{\gamma _\mu - \frac 4m p_\mu} u(p,s)
	\end{align}
	Hence, the structures containing $ \gamma _\mu $ or $ p_\mu $ include contributions from the $ 1/2 $ states. From Eq. \eqref{6.2}, it follows that only structure proportional to $ g_{\mu\nu} $ is free of $ 1/2 $ state contributions. 
	\item Not all Lorentz structures are independent. This problem can be solved by using the specific order of Dirac matrices. In the present work, we specify the desired order of Dirac matrices to be in the form $ \gamma _\mu \slashed \varepsilon \slashed q \slashed p \gamma _5 $. 
\end{enumerate} 
We choose the Lorentz structures $ \slashed \varepsilon \slashed p \gamma _5 q _\mu $, $ \slashed q \slashed p \gamma _5 \slashed \varepsilon q_\mu $, and $ \slashed q \slashed p \gamma _5 \varepsilon _\mu $ for the determination of the coupling constants $ g_1 $, $ g_2 $, and $ g_3 $ which are free from $ 1/2 $ contamination and which also yield better stability in the numerical analysis.
\par
The correlation function in the deep Euclidean domain, $ p^2 \ll 0 $ and $ (p+q)^2 \ll 0 $, can be calculated by using OPE near the light cone. The ample details of calculations are presented in \cite{Aliev2020} and for this reason, we do not repeat them here.
\par 
In the final step, performing a double Borel transformation over the variables $ -p^2 $ and $ -(p+q)^2 $, choosing the coefficients of the same Lorentz structures in both representations and matching them, and using the quark-hadron duality ans\"atz, we get the desired sum rules for these strong coupling constants:
\begin{gather}
	g_1 = -\frac 1{\lambda _1 \lambda _2 (m_1 + m_2)} e^{m_1^2/M_1^2 + m_2^2/M_2^2} \Pi _1^{(S)}\\
	g_2 = \frac 1{\lambda _1 \lambda _2} e^{m_1^2/M_1^2 + m_2^2 /M_2^2 } \Pi _2^{(S)}\\
	g_3 = \frac 1{\lambda _1 \lambda _2} e^{m_1^2/M_1^2 +m_2^2/M_2^2} \Pi _3^{(S)}
\end{gather}
While one discovers that all the terms vanish for the antisymmetric case, the explicit expressions of $ \Pi _i ^{(S)} $ can be found in \cite{Aliev2020}. 
\par
At the end of this section, we derive the corresponding coupling constants for the vertices $ B _{QQ}^* B _{QQ} \gamma $ by using the VMD ans\"atz. The VMD implies that the $ B _{QQ}^* B_{QQ}\gamma $ vertex can be obtained from $ B _{QQ}^* B_{QQ} V $ by converting the corresponding vector meson to a photon. From the gauge invariance, the $ B _{QQ}^* B _{QQ} \gamma $ vertex is parametrized similarly to the $ B _{QQ}^* B _{QQ} V $ vertex as follows:
\begin{align}
	\langle \gamma(q) B _{QQ}^*(p_2)  \vert B _{QQ}(p_1) \rangle &= \bar u ^\alpha (p_2) [
	g_1^\gamma (\varepsilon _\alpha^{*\gamma} \slashed q - q _\alpha \slashed \varepsilon^{*\gamma}) \gamma _5
	+ g_2^\gamma (P\cdot q \varepsilon _\alpha^{*\gamma} - P \cdot \varepsilon ^{*\gamma} q_\alpha)\gamma _5 
	\ar g_3^\gamma (q\cdot \varepsilon^{*\gamma} q^\alpha - q^2 \varepsilon _\alpha^{*\gamma}) \gamma _5		
	] u(p_1)
\end{align}
Obviously, the last term for real photons is equal to zero. To obtain the vertex $ B_{QQ}^* B_{QQ}\gamma $ from the $ B_{QQ}^*B_{QQ}V $, it is necessary to make the replacement
\begin{align}
	\varepsilon _\mu \to e \sum _{V=\rho^0, \omega, \phi} e_q\frac {f_V}{m_V}\varepsilon _\mu^\gamma \label{9}
\end{align}
and go from $ q^2 = m_V^2 $ to $ q^2 = 0 $. Let's check this statement.
\par
The radiative decays $ B _{QQ}^* \to B _{QQ} \gamma $ can be described by the following Lagrangian:
\begin{align}
	\lag = ie e_Q \bar Q \gamma _\mu Q A ^\mu + ie e_q \bar q \gamma_\mu q A^\mu
\end{align}
From this Lagrangian, one can obtain the decay amplitudes with the incorporation of the VMD, i.e.
\begin{align}
	\langle \gamma (q) B_{QQq}^*(p)  \vert \lag \vert B _{QQq} (p+q) \rangle &= 
		iee_q \varepsilon^{*\gamma\mu} \langle B _{QQq}^* (p) \vert \bar q \gamma _\mu q \vert B _{QQq} (p+q) \rangle
		\nn = ee_s \varepsilon  ^{*\gamma\mu}  \frac{\varepsilon _\mu}{q^2-m_\phi^2} \langle \phi (q) B_{QQs}^*(p) \vert B_{QQs} (p+q)\rangle 
		\nn + ee_q \varepsilon  ^{*\gamma\mu} \frac {\varepsilon _\mu}{q^2-m_\rho^2} \langle \rho (q) B _{QQq}^*(p) \vert B _{QQq} (p+q) \rangle
		\nn + ee_q \varepsilon ^{*\gamma\mu}  \frac {\varepsilon _\mu}{q^2 - m_\omega^2} \langle \omega (q) B _{QQq}^*(p) \vert B _{QQq} (p+q) \rangle
\end{align}
At the point $ q^2 = 0 $, (real photon case), this expression is simplified and we have
\begin{align}
	\langle \gamma (q)B _{QQq}^* (p) \vert \lag \vert B _{QQq} (p+q) \rangle &=
		ee_s \varepsilon^{*\gamma} \cdot \varepsilon \frac {f_\phi}{m_\phi} \langle \phi (q) B _{QQs}^*(p) \vert B _{QQs} (p+q) \rangle
		\nn + ee_q \varepsilon^{*\gamma} \cdot \varepsilon \frac {f_\rho}{m_\rho} \langle \rho (q) B _{QQq} ^*(p) \vert B _{QQq} (p+q) \rangle
		\nn + ee_q \varepsilon ^{*\gamma} \cdot \varepsilon \frac {f_\omega}{m_\omega} \langle \omega(q) B _{QQq}^*(p) \vert B _{QQq}(p+q) \rangle
\end{align}
From Eqs. \eqref{1} and \eqref{9}, for $ B^* _{QQq} B_{QQq}\gamma $ vertex, we get
\begin{align}
	\langle \gamma(q) B ^* _{QQq}(p)  \vert \lag \vert B _{QQq}(p+q) \rangle &= \sum _{V = \rho, \omega, \phi}ee_q \frac {f_V}{m_V} \bar u _\alpha (p)[
		g_1 (-q_\mu \slashed \varepsilon^{*\gamma} + \varepsilon _\mu^{*\gamma} \slashed q)
		\nn -g_2(P\cdot \varepsilon^{*\gamma} q_\mu - P\cdot q \varepsilon _\mu^{*\gamma})
	] \gamma _5 u(p+q) \label{10}
\end{align}
Comparing Eqs. \eqref{1} and \eqref{10}, we obtain the relation among the couplings $ B^*_{QQq} B _{QQq} V $ and $ B_{QQq}^* B_{QQq}\gamma $
\begin{align}
	g_i^\gamma = \begin{cases}
		e_s \frac{f_\phi}{m_\phi} g_i^\phi & \mbox{for }\Omega _{QQs}^* \to \Omega _{QQs}\gamma\\
		e_u \pp{\frac{f_\rho}{m_\rho} g_{i}^\rho + \frac{f_\omega}{m_\omega}g_{i}^\omega} & \mbox{for }\Xi^*_{QQu}\to \Xi _{QQu}\gamma\\
		e_d\pp{-\frac{f_\rho}{m_\rho}g_{i}^\rho+\frac{f_\omega}{m_\omega}g_{i}^\omega} & \mbox{for } \Xi ^*_{QQd} \to \Xi _{QQd}\gamma
	\end{cases}
\end{align}
for $ i=1,2 $. Here, we would like to make two remarks. First, we assume that couplings do not change considerably when we go from $ q^2 = m_V^2 $ to $ q^2 = 0 $. The second remark is related to the fact that, in principle, heavy vector meson resonances can also contribute. These contributions are neglected since in the heavy quark limit their contributions are proportional to $ m _{\rm heavy\ meson}^{-3/2} $.
\par 
In the numerical calculations for $ f_\rho $, $ f_\omega $, and $ f_\phi $, we have used the prediction of the sum rules $ f_\rho = 205 {\rm\ MeV} $, $ f_\omega = 185 {\rm\ MeV} $, and $ f_\phi = 215 {\rm\ MeV} $ \cite{Brown2014}. 
\par 
In this work, instead of the formfactors $ g_1^\gamma $ and $ g_2^\gamma $, we will use the magnetic dipole and electric quadrupole formfactors, $ G_M $ and $ G_E $, respectively, which are more convenient from an experimental point of view. The relation among these formfactors at the $ q^2=0 $ point are
\begin{gather}
	G_M = (3m_1+m_2) \frac{m_2}{3m_1} g_1^\gamma + (m_1-m_2) m_2 \frac{g_2^\gamma}3\\
	G_E = (m_1-m_2) \frac{m_2}{3m_1} (g_1^\gamma + m_1g_2^\gamma)
\end{gather}
Using these relations, it is straightforward to calculate the decay widths of $ B _{QQ}^* B_{QQ}\gamma $ decay.  The result is
\begin{align}
	\Gamma = \frac {3\alpha}4 \frac{k_\gamma^3}{m_2^2} (3G_E^2 +G_M^2) \label{24}
\end{align}
where $ \alpha $ is the fine structure coupling and $ k_\gamma = (m_1^2-m_2^2)/2m_1 $ is the photon energy.
\section{Numerical analysis}\label{sec:3}
In this section, we perform the numerical analysis of the LCSR for the coupling constants $ g_1 $ and $ g_2 $ obtained in the previous section for the $ \Xi _{QQ'}^*\Xi_{QQ'}\omega $ and $ \Omega _{QQ'}^*  \Omega _{QQ'} \phi $ vertices by using Package X \cite{Patel2015}.
\par 
The LCSR involves various input parameters, such as the quark masses, the masses and residues of doubly heavy baryons, and the decay constants of the light vector mesons, $ \omega $ and $ \phi $. These parameters are collected in Table \ref{tab:1.1}.
{
\begin{table}
	[H]\centering\setlength\tabcolsep{.2cm}
	\caption{Part of the input parameters. The masses and decay constants are at $ \mu = 1{\rm\ GeV} $ and in units of GeV.}\label{tab:1.1}
	\resizebox{\columnwidth}{!}{
	\begin{tabular}
		{cc|cc|cc|cc|cc|cc}
		\hline 
		\hline
		Parameter & Value & Parameter & Value & Parameter & Value & Parameter & Value & Parameter & Value & Parameter & Value\\
		\hline
		$ m_u $ & 0 & $ m_\omega $ & 0.783 & $ m _{\Xi _{cc}^*} $ & 3.692 \cite{Brown2014} & $ \lambda _{\Xi _{cc}^*} $ & 0.12 \cite{Aliev2013} & $ m _{\Xi _{cc}} $ & 3.610 \cite{Brown2014} & $ \lambda _{\Xi _{cc}} $ & 0.16 \cite{Aliev2012}\\
		$ m_d $ & 0 & $ f_\omega $ & 0.187 & $ m _{\Xi _{bb}^*} $ & 10.178 \cite{Brown2014} & $ \lambda _{\Xi _{cc}^*} $ & 0.22 \cite{Aliev2013}& $ m _{\Xi _{bb}} $ & 10.143 \cite{Brown2014} & $ \lambda _{\Xi _{bb}} $ & 0.44 \cite{Aliev2012}\\
		$ m_s $ & 0.137 & $ f_\omega^T $ & 0.151 & $ m _{\Xi _{bc}^*} $ & 6.985 \cite{Brown2014} & $ \lambda _{\Xi _{cc}^*} $ & 0.15 \cite{Aliev2013}& $ m _{\Xi _{bc}} $ & 6.943 \cite{Brown2014} & $ \lambda _{\Xi _{bc}} $ & 0.28 \cite{Aliev2012}\\
		$ m_c $ & 1.4 & $ m_\phi $ & 1.019 & $ m _{\Omega _{cc}^*} $ & 3.822 \cite{Brown2014} & $ \lambda _{\Omega _{cc}^*} $ & 0.14 \cite{Aliev2013} & $ m _{\Omega _{cc}} $ & 3.738 \cite{Brown2014} & $ \lambda _{\Omega _{cc}} $ & 0.18 \cite{Aliev2012}\\
		$ m_b $ & 4.8 & $ f_\phi $ & 0.215 & $ m _{\Omega _{bb}^*} $ & 10.308 \cite{Brown2014} & $ \lambda _{\Omega _{cc}^*} $ & 0.25 \cite{Aliev2013} & $ m _{\Omega _{bb}} $ & 10.273 \cite{Brown2014} & $ \lambda _{\Omega _{bb}} $ & 0.45 \cite{Aliev2012}\\
		& & $ f_\phi^T $ & 0.186 & $ m _{\Omega _{bc}^*} $ & 7.059 \cite{Brown2014} & $ \lambda _{\Omega _{cc}^*} $ & 0.17 \cite{Aliev2013} & $ m _{\Omega _{bc}} $ & 6.998 \cite{Brown2014} & $ \lambda _{\Omega _{bc}} $ & 0.29 \cite{Aliev2012}\\
		\hline 
		\hline 
	\end{tabular}
	}
\end{table}
}
\par
The main nonperturbative input parameters of the LCSR are the vector meson distribution amplitudes (DAs). The explicit expressions of the vector meson DAs are given in \cite{Aliev2020} and references therein. The parameters that appear in the light vector meson DAs for $ \omega $ and $ \phi $ are presented in Table \ref{tab:1.2}. 
\begin{table}
	[H]\centering\setlength\tabcolsep{.5cm}
	\caption{Vector meson DA parameters for $ \omega $ and $ \phi $ at $ \mu = 1 {\rm\ GeV} $ \cite{Ball1998, Ball1999, Ball1996, Ball2006, Ball19992, Ball2005}. The accuracy of these parameters are 30--50\%.}\label{tab:1.2}
	%\resizebox{\columnwidth}{!}{
	\begin{tabular}
		{ccc|ccc}
		\hline 
		\hline
		Parameter & $ \omega $ & $ \phi $ & Parameter & $ \omega $ & $ \phi $\\
		\hline 
		$ a_1^\parallel $ & 0 & 0 & $ \kappa _3^\perp $ & 0 & 0\\
		$ a_1 ^\perp $ & 0 & 0 & $ \omega_3^\perp $ & 0.55 & 0.20\\
		$ a_2^\parallel $ & 0.15 & 0.18 & $ \lambda_3^\perp $ & 0 & 0\\
		$ a_2^\perp $ & 0.14 & 0.14 & $ \zeta_4^\parallel $ & 0.07 & 0\\
		$ \zeta _3 ^\parallel $ & 0.030 & 0.024 & $ \tilde \omega _4 ^\parallel $ & --0.03 & --0.02\\
		$ \tilde \lambda _3 ^\parallel $ & 0 & 0 & $ \zeta _4 ^\perp $ & --0.03 & --0.01\\
		$ \tilde \omega _3 ^\parallel $ & --0.09  & --0.045 & $ \tilde \zeta _4 ^\perp $ & --0.08 & --0.03 \\
		$ \kappa_3^\parallel $ & 0 & 0 & $ \kappa_4^\parallel $ & 0 & 0\\
		$ \omega _3 ^\parallel $ & 0.15 & 0.09 & $ \kappa_4^\perp $ & 0 & 0\\
		$ \lambda_3^\parallel $ & 0 & 0 \\
		\hline
		\hline
	\end{tabular}
	%}
\end{table}
The LCSR for the strong coupling constants $ g_1 $ and $ g_2 $ involves three auxiliary parameters, namely the Borel mass parameter, $ M^2 $, the continuum threshold $ s_0 $, and the parameter $ \beta $, in the expression of the interpolating current. Hence, we need to find the working regions of these parameters where the results for the coupling constants $ g_1 $ and $ g_2 $ practically exhibit insensitivity to the variation of these parameters. The lower bound of $ M^2 $ is determined by requiring the contributions of higher twist terms considerably small than the leading twist one (say than 15\%). The upper bound of $ M^2 $ can be found by requiring that the continuum contribution to the sum rules should be less than 25\% of the total result. The value of continuum threshold $ s_0 $ is obtained by demanding that the two-point sum rules reproduce the mass of doubly heavy baryons with 10\% accuracy. After performing the numerical analysis, we obtained the working regions for $ M^2 $ and $ s_0 $ as displayed in Table \ref{tab:2}.
\begin{table}
	[H]\centering\setlength\tabcolsep{.5cm}
	\caption{The working regions of the Borel mass parameter and the central value of the continuum threshold.}\label{tab:2}
	\begin{tabular}
		{ccc}
		\hline 
		\hline 
		Transition &  $ M^2 \ ({\rm GeV^2}) $ & $ s_0 \ ({\rm GeV^2}) $\\
		\hline 
		$ \Xi _{cc}^* \to \Xi _{cc} \omega $ & $ 3 \leq M^2 \leq 4.5 $ & 18\\
		$ \Xi _{bb}^* \to \Xi _{bb} \omega $ & $ 8\leq M^2 \leq 12 $& 110\\
		$ \Xi _{bc}^* \to \Xi _{bc} \omega $ & $ 6\leq M^2 \leq 8 $& 60\\ \hline
		$ \Omega _{cc}^* \to \Omega _{cc} \phi $ & $ 3\leq M^2 \leq 5 $ & 18\\
		$ \Omega _{bb}^* \to \Omega _{bb} \phi $ & $ 8 \leq M^2 \leq 13 $& 110\\
		$ \Omega _{bc}^* \to \Omega _{bc} \phi $ & $ 6\leq M^2 \leq 9 $ & 60\\
		\hline 
		\hline
	\end{tabular}
\end{table}
Finally, we note that the value of the $ \Xi _{QQ}^* \to \Xi _{QQ}\rho^0 $ couplings can be obtained from the results of \cite{Aliev2020} via the isospin symmetry. 
\par
\begin{figure}
	[H]\centering
	\includegraphics[width=.8\textwidth]{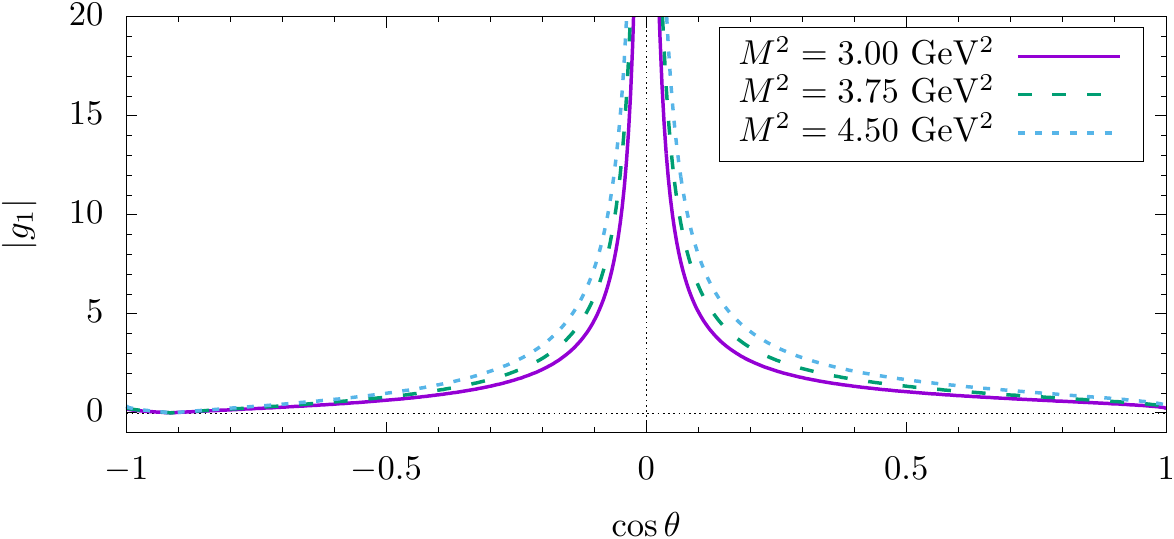}
	\caption{The dependence of the modulus of the coupling constant $ g_1 $ for $ \Xi _{cc}^* \to \Xi _{cc}\omega $ on $ \cos\theta $ at the shown values of $ M^2 $ with $ s_0 = 18 {\rm\ GeV^2} $.}\label{fig:1}
\end{figure}
\begin{figure}
	[H]\centering
	\includegraphics[width=.8\textwidth]{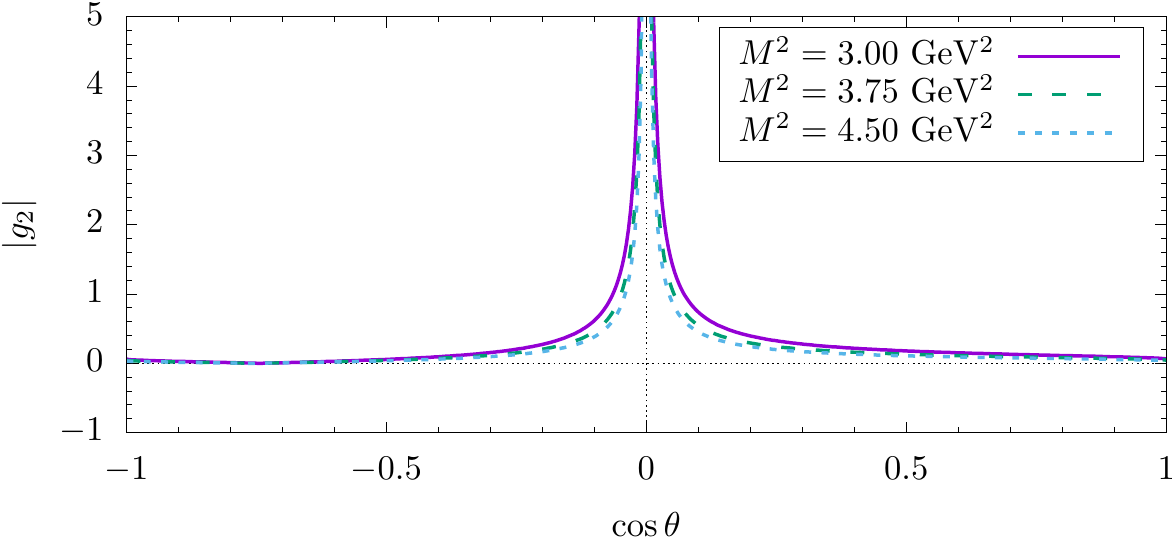}
	\caption{The same as Fig. \ref{fig:1} but for $ g_2 $.}\label{fig:2}
\end{figure}
\begin{figure}
	[H]\centering
	\includegraphics[width=.8\textwidth]{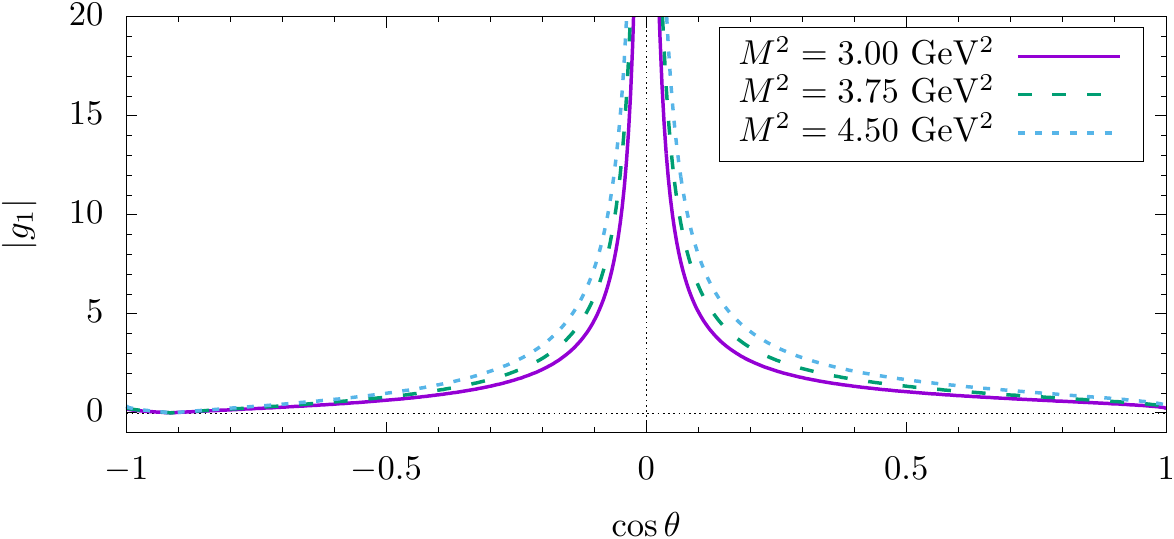}
	\caption{The same as Fig. \ref{fig:1} but for $ g_3 $.}\label{fig:3}
\end{figure}
\begin{figure}
	[H]\centering
	\includegraphics[width=.8\textwidth]{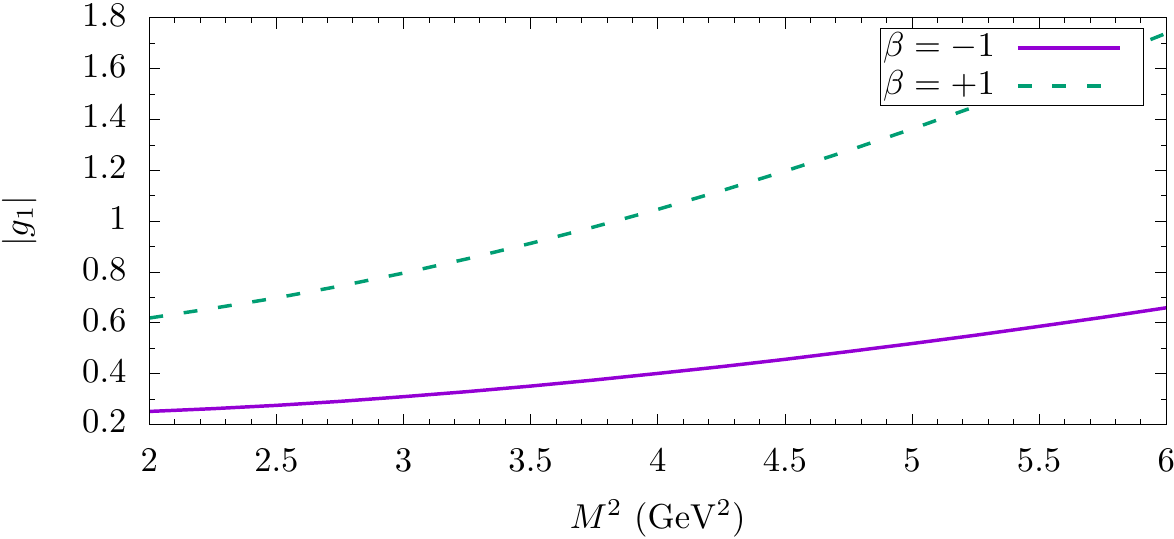}
	\caption{The dependence of the modulus of the coupling constant $ g_1 $ for $ \Xi _{cc}^* \to \Xi _{cc} \omega $ on $ M^2 $ at the shown values of $ \beta $ with $ s_0 = 18 {\rm\ GeV^2} $.}\label{fig:4}
\end{figure}
\begin{figure}
	[H]\centering
	\includegraphics[width=.8\textwidth]{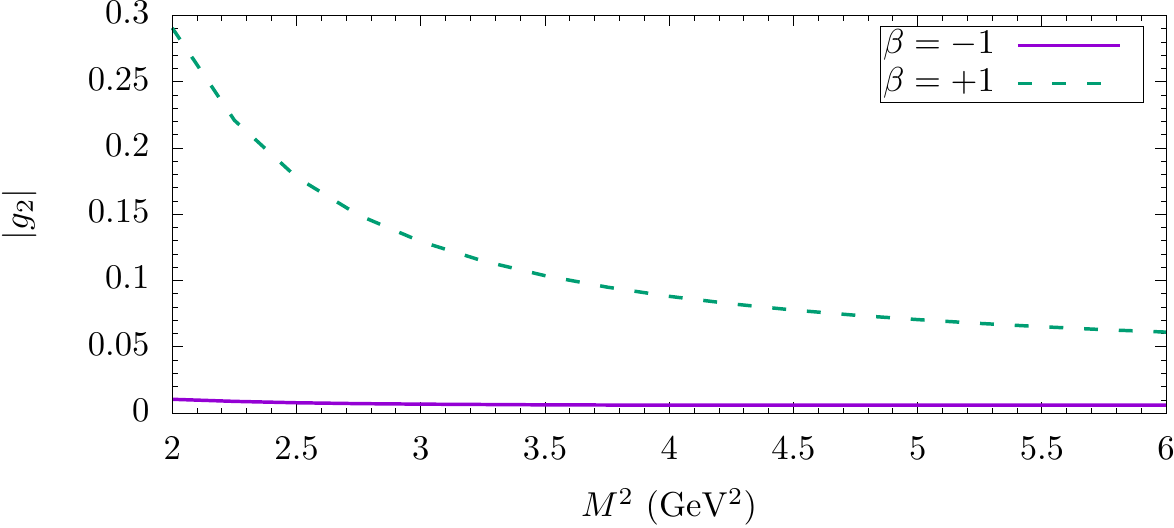}
	\caption{The same as Fig. \ref{fig:4} but for $ g_2 $.}\label{fig:5}
\end{figure}
\begin{figure}
	[H]\centering
	\includegraphics[width=.8\textwidth]{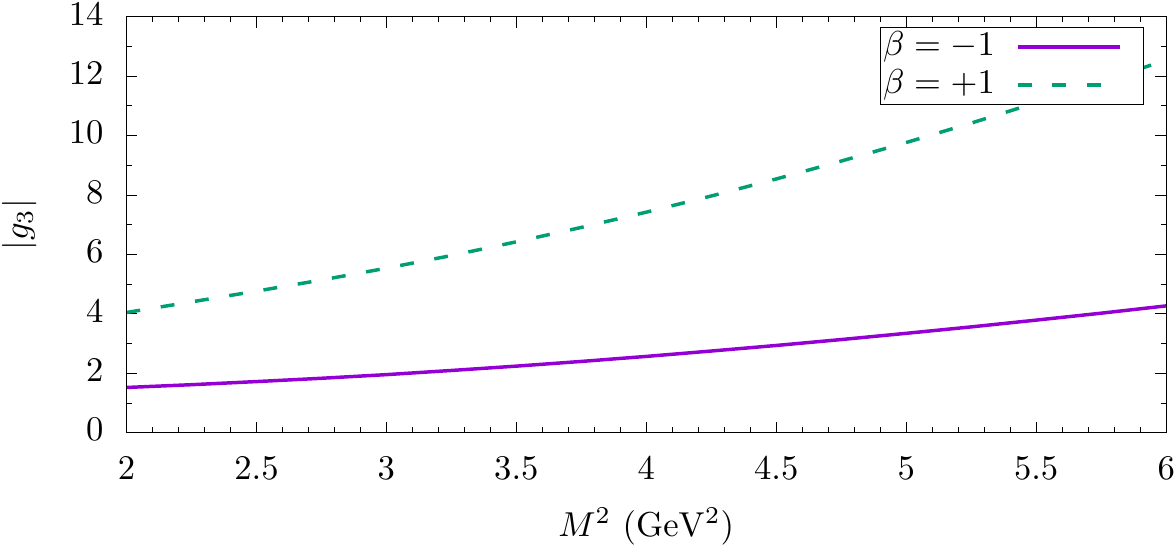}
	\caption{The same as Fig. \ref{fig:4} but for $ g_3 $.}\label{fig:6}
\end{figure}

As an illustration, we present the dependence of the coupling constants $ g_1 $, $ g_2 $, and $ g_3 $ on $ \cos\theta $ for the transition $ \Xi _{cc}^* \to \Xi _{cc}\omega $, where $ \theta $ is defined via $ \beta = \tan \theta $ and on the Borel mass parameter, $ M^2 $ in Figs. \ref{fig:1}--\ref{fig:6}. We summarized our results in Table \ref{tab:4}. The corresponding values for the case of the Ioffe current, for which $ \beta = -1 $, are also presented. One can see that in Figs. \ref{fig:1}--\ref{fig:3}, the value of the coupling constant practically does not change for the values of $ \vv{\cos\theta} $ between 0.5 and 0.8, hence we determine the working region of $ \beta $ accordingly. The errors in Table \ref{tab:4} reflect the uncertainties in the aforementioned input parameters. From this table, it follows that in the case of a general current, the values of the coupling constants are comparable to those in the case of the Ioffe current.
\begin{table}
	[H]\centering\setlength\tabcolsep{.5cm}
	\caption{The obtained values of the moduli of the coupling constants $ g_1 $, $ g_2 $, and $ g_3 $ for the aforementioned transitions accompanied by a light vector meson.}\label{tab:4}
	\resizebox\columnwidth!{\begin{tabular}
		{c|ccc|ccc}
		 \multicolumn1c{}& \multicolumn{3}{c}{Case of the general current} & \multicolumn{3}{c}{Case of the Ioffe current}\\
		\hline
		\hline
		Transition & $ \vv{g_1} $ & $ \vv{g_2} $ & $ \vv{g_3} $ & $ \vv{g_1} $ & $ \vv{g_2} $ & $ \vv{g_3} $\\
		\hline
		$\Xi_{cc}^*\to\Xi_{cc}\rho^0$	&$	1.13\pm0.25	$&$	0.11\pm0.03	$&$	7.81\pm1.83	$&$	0.99\pm0.22	$&$	0.10\pm0.02	$&$	6.92\pm1.62	$ \\
		$\Xi_{bb}^*\to\Xi_{bb}\rho^0$	&$	0.76\pm0.23	$&$	0.03\pm0.00	$&$	15.19\pm4.64	$&$	0.67\pm0.20	$&$	0.02\pm0.00	$&$	13.45\pm4.11	$ \\
		$\Xi_{bc}^*\to\Xi_{bc}\rho^0$	&$	1.06\pm0.20	$&$	0.05\pm0.01	$&$	14.44\pm2.84	$&$	0.94\pm0.18	$&$	0.05\pm0.01	$&$	12.79\pm2.51	$ \\ \hline
		$\Xi_{cc}^*\to\Xi_{cc}\omega$	&$	1.02\pm0.23	$&$	0.10\pm0.02	$&$	7.10\pm1.68	$&$	0.90\pm0.20	$&$	0.09\pm0.02	$&$	6.29\pm1.49	$ \\
		$\Xi_{bb}^*\to\Xi_{bb}\omega$	&$	0.69\pm0.21	$&$	0.03\pm0.00	$&$	13.82\pm4.25	$&$	0.61\pm0.19	$&$	0.02\pm0.00	$&$	12.24\pm3.77	$ \\
		$\Xi_{bc}^*\to\Xi_{bc}\omega$	&$	0.97\pm0.19	$&$	0.05\pm0.01	$&$	13.14\pm2.60	$&$	0.86\pm0.17	$&$	0.05\pm0.01	$&$	11.64\pm2.31	$ \\ \hline
		$\Omega_{cc}^*\to\Omega_{cc}\phi$	&$	1.50\pm0.32	$&$	0.50\pm0.14	$&$	9.79\pm2.50	$&$	1.32\pm0.28	$&$	0.45\pm0.13	$&$	8.64\pm2.22	$ \\
		$\Omega_{bb}^*\to\Omega_{bb}\phi$	&$	1.22\pm0.35	$&$	0.15\pm0.03	$&$	23.90\pm7.19	$&$	1.08\pm0.31	$&$	0.14\pm0.03	$&$	21.15\pm6.38	$ \\
		$\Omega_{bc}^*\to\Omega_{bc}\phi$	&$	1.47\pm0.29	$&$	0.25\pm0.04	$&$	19.26\pm4.13	$&$	1.30\pm0.26	$&$	0.22\pm0.03	$&$	17.03\pm3.66	$ \\
		\hline 
		\hline
	\end{tabular}}
\end{table}
\par
Now using the obtained results for $ g_1 $ and $ g_2 $, we can estimate $ g_i^\gamma $ and hence $ G_M $ and $ G_E $. The results for $ G_M $ and $ G_E $ are collected in Table \ref{tab:5}. 
\begin{table}
	[H]\centering\setlength\tabcolsep{.5cm}
	\caption{The electric quadrupole and magnetic dipole formfactors for the shown transitions.}\label{tab:5} 
	\begin{tabular}
		{ccc}
		\hline
		\hline
		Transition & $ \vv{G _E} $ & $ \vv{G _M} $\\
		\hline
		$\Xi_{cc}^{*++}\to\Xi_{cc}^{++}\gamma$	&	$0.00\pm0.00$	&	$1.78\pm0.40$	\\
		$\Xi_{cc}^{*+}\to\Xi_{cc}^{+}\gamma$	&	$0.00\pm0.00$	&	$0.11\pm0.02$	\\
		$\Xi_{bb}^{*0}\to\Xi_{bb}^{0}\gamma$	&	$0.00\pm0.00$	&	$3.41\pm1.03$	\\
		$\Xi_{bb}^{*-}\to\Xi_{bb}^{-}\gamma$	&	$0.00\pm0.00$	&	$0.22\pm0.06$	\\ \hline
		$\Omega_{cc}^{*+}\to\Omega_{cc}^{+}\gamma$	&	$0.00\pm0.00$	&	$0.52\pm0.11$	\\
		$\Omega_{bb}^{*-}\to\Omega_{bb}^{-}\gamma$	&	$0.00\pm0.00$	&	$1.18\pm0.34$	\\
		\hline
		\hline
	\end{tabular}
\end{table}
\par
Using Eq. \eqref{24} and the values of $ G_M $ and $ G_E $ for the decay widths of these transitions, it is straightforward to find the values of the corresponding decay widths. From Eq. \eqref{24}, one can see that the decay width is very sensitive to the mass difference of the considered baryons, $ \Delta m = m_1-m_2 $. Therefore, a tiny change in the mass difference leads to a significant change in the decay width. To see this, as an example, we present the decay widths for the transition $ \Omega _{ccs}^* \to \Omega _{ccs}\gamma $ by using the different mass differences obtained in various approaches. The results are presented in Table \ref{tab:5.1}. 
\begin{table}
	[H]\centering\setlength\tabcolsep{.2cm}
	\caption{The decay width of the transition $ \Omega _{ccs}^*\to \Omega _{ccs}\gamma $ for different mass splittings.}\label{tab:5.1}
	\begin{tabular}
		{c|ccccccc}
		\hline
		\hline
		$ \Delta m $ [MeV] & 57 \cite{Lu2017} & 61 \cite{Bernotas2013} & 73 \cite{Hackman1978} & 84 \cite{Brown2014} & 94 \cite{Branz2010,Xiao2017,Cui2018} & 100 \cite{Li2017b} \\
		\hline
		$ \Gamma $ [keV] & 0.07 & 0.09 & 0.15 & 0.23 & 0.33 & 0.40 \\
		\hline 
		\hline 
	\end{tabular}
\end{table}
In our numerical calculations, for the masses of spin-1/2 and spin-3/2 states, we have used the results of \cite{Brown2014} (see Table \ref{tab:1.1}) because the results are practically free from errors. Our final results on the decay widths are collected in Table \ref{tab:6}. For completeness, we also presented the results for corresponding decay widths obtained within different approaches. From the comparison of decay widths, we see that our result only for the $\Omega _{cc}^* \to \Omega _{cc}\gamma $ decay is close to the prediction of the lattice theory and considerably different from the ones in other existing approaches. One possible source of these discrepancies may be that, for doubly heavy baryon systems, the VMD ans\"atz may work not so quite well. In order to see how the VMD works for doubly heavy baryon systems, it would be useful to calculate $G_M$ and $G_E$ directly, i.e. without using the VMD ans\"atz. This work is in progress.
\begin{table}
	[H]\centering\setlength\tabcolsep{.1cm}
	\caption{The widths of the shown radiative decays in units of keV.}\label{tab:6}
	\resizebox\columnwidth!{\begin{tabular}
		{cccccc}
		\hline 
		\hline
		Transition & Our work & Chiral quark model \cite{Xiao2017} & Three-quark model \cite{Branz2010} & Chiral perturbation theory \cite{Li2017b} & Lattice QCD \cite{Bahtiyar2018} \\
		\hline
		$ \Xi _{cc}^{*++}\to \Xi _{cc}^{++}\gamma $ & $ (71.33\pm3.56)\times 10^{-2} $ & 16.7 & 23.5 & 22 & $ 7.77 \times 10^{-2} $\\
		$ \Xi _{cc}^{*+} \to \Xi _{cc}^+\gamma $ & $ (0.29\pm0.01)\times 10^{-2} $ & 14.6 & 28.8 & 9.57 & $ 9.72 \times 10^{-2} $\\
		$ \Omega _{cc}^* \to \Omega _{cc}\gamma $ & $ (6.08\pm0.28)\times 10^{-2} $ & 6.93 & 2.11 & 9.45 & $ 8.47 \times 10^{-2} $\\
		$ \Xi _{bb}^{*0} \to \Xi _{bb}^0\gamma $ & $ (2.64\pm0.24)\times 10^{-2} $ & 1.19 & 0.31 & -- & --\\
		$ \Xi _{bb}^{*-} \to \Xi _{bb}^-\gamma $ & $ (0.01\pm0.00)\times 10^{-2} $ & 0.24 & 0.06 & -- & --\\
		$ \Omega _{bb}^* \to \Omega _{bb}\gamma $ & $ (0.31\pm0.03)\times 10^{-2} $ & 0.08 & 0.02 & -- & --\\
		\hline
		\hline
	\end{tabular}}
\end{table}
\section{Conclusion}\label{sec:4}
In the present work, first, we estimated the strong coupling constants of $ B _{QQ'}^* B _{QQ'} V $ vertices within the framework of the LCSR method. Then, assuming the VMD model, we calculated the magnetic dipole and electric quadrupole formfactors, $ G_M $ and $ G_E $, respectively, at the point $ Q^2=0 $. Using the results for $ G_M $ and $ G_E $, we obtained the decay widths of the radiative decays $ B_{QQ}^* \to B _{QQ} \gamma $. Our result for the decay widths of $\Omega_{cc}^*\to\Omega_{cc}\gamma$ is in good agreement with the lattice result and considerably different from the prediction of other channels in various approaches. Our predictions on the strong coupling constants for the radiative $ B _{QQ}^* B _{QQ}V  $ vertices, as well as the decay widths, can be checked at LHCb experiments in the future.
%\appendix
%\section{}\label{app:A}
%\section{}\label{app:B}
\bibliography{refs}
\end{document}